# Wave propagation and energy dissipation in collagen molecules

Mario Milazzo [†,‡], Gang Seob Jung [†], Serena Danti [†, ‡, §], Markus J. Buehler [†,*]

[†]Laboratory for Atomistic and Molecular Mechanics (LAMM), Dept. of Civil and Environmental Engineering, Massachusetts Institute of Technology, United Stated of America

[‡]The BioRobotics Institute, Scuola Superiore Sant'Anna, Italy

[§]Dept. of Civil and Industrial Engineering, University of Pisa, Italy

**\* Correspondence to Markus J. Buehler:** Laboratory for Atomistic and Molecular Mechanics (LAMM), Department of Civil and Environmental Engineering, Massachusetts Institute of Technology, 77 Massachusetts Ave., Cambridge, Massachusetts 02139 United States of America. mbuehler@mit.edu

**Abstract**

Collagen is the key protein of connective tissue (i.e., skin, tendons and ligaments, cartilage, among others) accounting for 25% to 35% of the whole-body protein content, and entitled of conferring mechanical stability. This protein is also a fundamental building block of bone due to its excellent mechanical properties together with carbonated hydroxyapatite minerals. While the mechanical resilience and viscoelasticity have been studied both *in vitro* and *in vivo* from the molecule to tissue level, wave propagation properties and energy dissipation have not yet been deeply explored, in spite of being crucial to understand the vibration dynamics of collagenous structures (e.g., eardrum, cochlear membranes) upon impulsive loads. By using a bottom-up atomistic modelling approach, here we study a collagen peptide under two distinct impulsive displacement loads, including longitudinal and transversal inputs. Using a one-dimensional string model as a model system, we investigate the roles of hydration and load direction on wave propagation along the collagen peptide and the related energy dissipation. We find that wave transmission and energy-dissipation strongly depend on the loading direction. Also, the hydrated collagen peptide can dissipate five times more energy than dehydrated one. Our work suggests a distinct role of collagen in term of wave transmission of different tissues such as tendon and eardrum. This study can step towards understanding the mechanical behaviour of collagen upon transient loads, impact loading and fatigue, and designing biomimetic and bio-inspired materials to replace specific native tissues such as the tympanic membrane.

# 1. Introduction

Collagen is a natural polymer richly abundant in human and animal tissues, such as in tendon, ligament, cartilage, skin, and bone.[1–4] Based on the molecular structure, researchers have classified 28 different collagen types. Among them, the type I is the most abundant collagen in the human body.[1,5,6] In bone tissue, collagen type I constructs with hydroxyapatite an interesting hierarchical structure that allows stability and elasticity for the musculo-skeletal system.[1,7–15] From macroscopic collagenous tissue (e.g., tendon) with dimensions on the order of centimeters to the sequence of amino acids, it is possible to recognize specific structures at distinct scales of observation (Figure 1A). Collagen fibers of the macro tissues are composed of collagen fibrils that result from a staggered parallel organization of multiple polypeptides, known as tropocollagen, with a periodicity universally known as D-band, in which D is a characteristic dimension of 67 nm. A gap has also been observed between two adjacent molecules measuring 0.54D ($\approx$ 36 nm),[16] which is filled with a mineral component in the building block of bone tissue.[17] Tropocollagen, the molecular building block, is a repeating sequence of the GXY triplet, in which G is glycine and X and Y are commonly proline and hydroxyproline.[18,19] The three-polypeptide chains form a triple-helix structure of the tropocollagen with length of $\approx$300 nm and a diameter of $\approx$1.5 nm. Earlier studies have extensively investigated collagen mechanics at different scales in terms of strength, toughness, loading-rate dependency, and viscoelasticity[20,21] based on experimental campaigns employing X-ray diffraction techniques[22–24] complemented with molecular modeling.[25–28] These mechanical properties imply topology variations in time scales that may range from microseconds to hours.[29]

An important advancement has recently been delivered by a number of studies investigating the role of water in collagen structure and properties, especially its mechanical features. Although dry collagen is critical for applications such as fabrication of leather or parchment,[30,31] hydrated collagen is its native state[32–35] and its stiffness is largely smaller than the dehydrated one due to the massive presence of protein-solvent hydrogen bonds (H-bonds) that prevent the formation of intramolecular H-bonds, mainly responsible for the increase of backbone rigidity.[21,36]

While previous studies have extensively presented the quasi-static properties of collagen,[21,37,38] little effort has been made to understand the capability of this material to transmit and dissipate mechanical energy along the structure when subjected to impulsive loads. A preliminary study reported on the propagation of a longitudinal force wave aiming at observing the delay in the crosslinks. However, a more extensive discussion on the energy dissipation along the structure is missing. Earlier work has successfully compared a theoretical approach with simulations to describe longitudinal wave propagation that is in the order of km·s$^{-1}$, and a time delay (~ns) that depends on the length of the triple helix.[39] The dynamic properties of H-bonds also play an interesting role in governing energy transfer in other proteins such as amyloid and silk, which stems from the unique properties of beta sheet crystals. Some relevant work by Xu *et al.* focused on material responses upon impulsive displacements.[40,41]

Experimentally, the scientific question, to define the different role of collagen in the different tissue, is difficult to address due the specific equipment requirements that must provide excellent time resolutions and reliable handling for positioning the necessary probes.[42,43] In this work, we perform an *in silico* study on dry and hydrated collagen peptides aiming at unveiling how energy is transferred and dissipated along the triple helix when a longitudinal or transversal impulsive load is applied to the free end of the structure (Figure 1B).

The insights hereby discussed can open new avenues of research in different fields such as biology, materials science, and tissue engineering. In view of this, a very interesting application is represented by the eardrum, a thin concave membrane in the middle ear, devoted to collect, filter, and transfer acoustic waves towards the inner ear. Its structure, is mainly composed of collagen type II fibers arranged in radial, circumferential, parabolic patterns. This specific structure, in addition to creating a scaffold to bear acoustic loads, plays a key role for the mechanical response and linking molecular to vibrational responses across scales.[40,44–49] There exists a specific orientation of such collagenous fibers, and thus of their building blocks, associated to the hydration state might significantly affect the wave transmission and dissipation, enabling the observed transversal vibrations. More broadly, understanding the dissipative behavior of tropocollagen opens not only to a deeper knowledge of materials biology but also to the optimized design of new biomimetic and bio-inspired materials able to deliver high-quality performances in tissue engineering applications with transient loads such as fatigue, impacts, or acoustical properties and mechanisms.

## 2. Materials and Methods

### 2.1. MD model preparation and relaxation

We prepare a (GPO)$_{20}$ peptide topology with the Triple-Helical collagen Building Script (THe BuScr) with length of about 180 Å.[50] In order to investigate the role of the water in the mechanical energy transfer, in addition to a dry structure (DS), we solvate the box that has the dimensions of 40 × 6 × 6 nm and randomly ionize (0.5 mol/L) the peptide via Visual Molecular Dynamics, reaching a hydrated structure (HS).

The equilibration of both the DS and HS is performed via LAMMPS using the CHARMM force field that includes the hydroxyproline residue.[51] We use the particle-particle particle-mesh solver with 10$^{-4}$ kcal/mol-Angstrom accuracy to compute the long-range Coulombic interactions, while for the short-range ones we employ the Leonard-Jones potential with global switching cut-offs set to 1 nm and 1.2 nm. The systems are equilibrated aiming at minimizing the potential energy at 310 K (37 °C), reaching a convergence of the root mean square deviation after 30 ns, according also to previous works.[21]

### 2.2. Wave propagation

We use a Molecular Dynamics (MD) approach to study the stress wavefront tracking to provide the main kinematic parameters to investigate the energy transfer along a collagen peptide.[41,52] We fix three backbone nitrogen atoms (BB-N) at the extremity of the three chains of the peptide and load the structures through the other three BB-N atoms at the opposite edge.

Two different load cases for both the DS and HS are here investigated: longitudinal (Longitudinal Case – LC) and transversal (Transversal Case – TC) cases due to the differences in the topology along and perpendicularly to the triple-helix structure (Figure 2). In the LC, the peptide is loaded with an impulsive displacement ($\Delta$) equal to 10 Å, as depicted in Figure 2A and Figure 2C, with a slope of 100 m/s[41] in a total time ($T$) equals 10 ps. Concerning the TC, the peptide is firstly pre-stretched axially with a slope of 100 m/s up to a fixed deformation (i.e., from 1 to 10% axial strain to contain the computational cost)[27] and, after 10 ps, we load the peptide edge with a vertical impulsive displacement, similar to the one used for the LC (Figure 2B and Figure 2D). For both the LC and TC, the overall observation time is 160 ps.

Wave speeds are estimated through the displacements in time of the backbone alpha Carbon atoms ($C_\alpha$) while the mechanical energy dissipation is investigated through the relaxation time ($\tau$), defined as a coefficient of the exponential function $\left(e^{-\frac{t}{\tau}}\right)$ that fits the square of the maximum wave speed as a function of time ($v_{Max}^2(t)$). In order to further analyze the wave speed results, we model the peptide for the LC as a rod and the TC as a vibrating string. These assumptions allow us to study the problem with analytical references and correlate the kinematic outcomes with the Young's modulus ($E$) according to the following equations:

$$v_{LC} = \sqrt{\frac{E}{\rho}} \qquad v_{TC} = \sqrt{\frac{E}{\rho}\varepsilon} \qquad \text{Eq. 1}$$

in which $v_{LC}$ and $v_{TC}$ are the wave speeds in the LC and TC respectively, whilst $\rho$ and $\varepsilon$ are the peptide density and tensile strain respectively.[53] To make a clear comparison between cases and structures, we set the maximum displacement in Figure 3 and Figure 4 to 15 Å. Finally, we estimate the specific acoustic impedance $Z$ through its definition:

$$Z = \rho v \qquad \text{Eq. 2}$$

in which $v$ is the wave speed.

## 3. Results and discussion

We perform a study on energy transmission and dissipation on two collagen peptides: one dry (Dry Structure – DS) and one hydrated (Hydrated Structure – HS) exposed to loads longitudinally (Longitudinal Case – LC – Figure 2A,C) or transversally (Transversal Case – TC – Figure 2B,D) to the triple-helix axis. As for this latter case, due to the similarities with a vibrating string, we pre-stretch the structure up to a tensile strain of 10% in order to reduce the computational costs. In both cases, the load consists of an impulsive displacement able to deliver a travelling displacement wave.

Figure 3 reports the results for the LC in which the axial displacement of the DS (Figure 3A) and the HS (Figure 3B) are plotted over time and position along the peptide structure. For both structures, the axial displacement-based load is almost completely dissipated before the induced wave reaches the fixed edge. In addition to the plots, the energy dissipation is estimated through the relaxation time ($\tau_{LC}$) that is comparable for both the DS and HS in the order of 100 ps. We believe that the water environments for the HS case does not play a significant role for longitudinal loads where axial stress waves are directly channelled into heat.[53] In contrast, measured wave speeds ($v_{LC}$) show a significant difference between the DS and HS, being higher for the dry peptide (i.e., 3082 m·s$^{-1}$ vs. 2190 m·s$^{-1}$). Compared to the fibrous beta-sheet-rich proteins investigated in [41] through a similar approach, wave speeds along the collagen peptides are definitely smaller. This result may be attributed to the almost crystalline structure of the proteins studied by the authors that is much stiffer of about one order of magnitude. Moreover, these results show velocities that are much slower than the wave speed possessed by the bone exposed to similar conditions (i.e., 4080 m·s$^{-1}$).[54] We attribute this deviation mainly to the absence of the mineral component, similar to carbonate hydroxyapatite, that is able to stiffen the whole topology and, thus, to enhance wave propagation according to 

$$vLC=\sqrt{\frac{E}{\rho}} \qquad v_{TC}=\sqrt{\frac{E}{\rho}}\varepsilon$$

Eq. 1. In view of possible application for the mechanics of hearing, we also use $Z=\rho v$ Eq. 2 to estimate the acoustic impedances that result in 2.63×10$^6$ kg·m$^{-2}$·s$^{-1}$ and 1.87×10$^6$ kg·m$^{-2}$·s$^{-1}$ for the DS and HS, respectively that results in, also, less than the bone's (7.75×10$^6$ kg·m$^{-2}$·s$^{-1}$).[54]

The estimated Young's Moduli for the DS and HS are 8.05 GPa and 4.07 GPa, respectively; while the result for the HS is quite similar to the one obtained in [55] with an error of 2%, the stiffness of the DS is slightly lower than what it is obtained in [21] (i.e., 10 MPa). The differences mainly depend on the method used to evaluate the mechanical properties that, for [21,55], relies on a tensile test on the structure; in contrast, we achieve our results from the wave speed propagation in which the loading rate plays a key role in the estimation of the mechanical properties,[27] an effect that is negligible in [41] because of the quasi-crystallinity of the analyzed structures. Furthermore, we employ 

$$vLC=\sqrt{\frac{E}{\rho}} \qquad v_{TC}=\sqrt{\frac{E}{\rho}}\varepsilon$$

Eq. 1 to correlate the wave speed with the Young's modulus but it represents only an approximation since it describes the longitudinal phenomena at the macro scale without taking into account other three-dimensional secondary effects at the molecular scale that induce transversal deformation due to the reduced bending stiffness of the molecule. The three-dimensionality of the structure deformation can be, thus, related to the errors in estimating the Young's modulus.[39]

Results for the TC, schematically reported in Figure 4, show a different perspective. The DS is able to almost fully convey the energy content induced by the external load with a reduced dissipation (average $\tau_{DS\_TC}$ = 429 ns). As shown in Figure 4A,C, the DS presents a partial reflection of the wave with a dampening effect at the free end that is due to the free degrees of freedom of the peptide. In contrast, for the HS, when the structure is exposed to a transversal load, the wave propagation is quickly annihilated (average $\tau_{HS\_TC}$ = 78 ns ≈ 0.2·$\tau_{WS\_TC}$). Interestingly, these dissipating phenomena differ considerably from what we described for the LC. Specifically, while we observe similar behaviors in the LC for the DS and HS, the perpendicularity between the loading and propagation directions in the TC, enhances the role of water, and specifically its viscosity, in damping the energy. However, although we observe a reduced damping for the DS in the TC, we believe that structural features of collagenous tissues, which are not modeled in the present work, might exist to compensate the weak dampening behavior. Quantitatively, a paired comparison remarks a four time increase of the relaxation time for the DS and a slight reduction for HS. Concerning the wave propagation for TC, it is possible to appreciate in

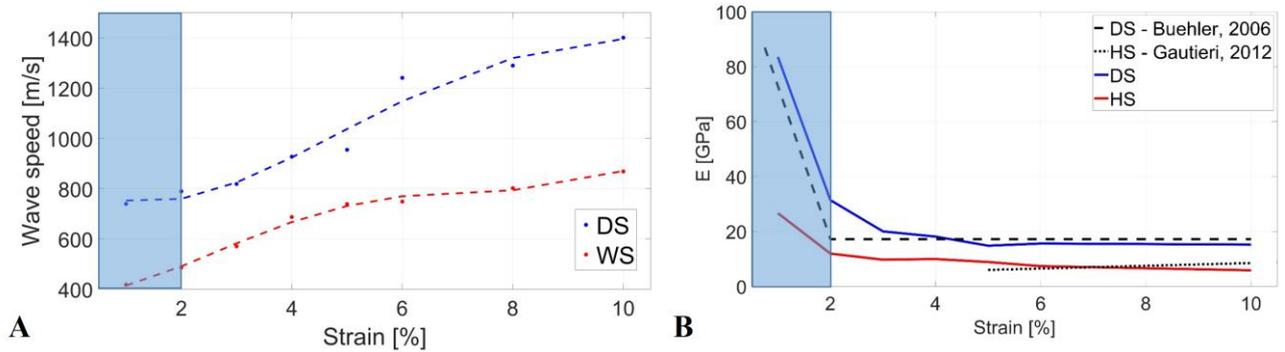

Figure 5 how the outcomes clearly confirm the theory of the vibrating string expressed also by $vLC = \sqrt{\frac{E}{\rho}}$

$$v_{TC} = \sqrt{\frac{E}{\rho}}\varepsilon$$     Eq. 1. In both the DS and HS the slope of the

travelling displacement (i.e., $v_{TC}$) gets steeper when the pre-strain increases and, thus, the Young's modulus.

To better compare the capability of the two structures in transferring the kinetic energy, we report in

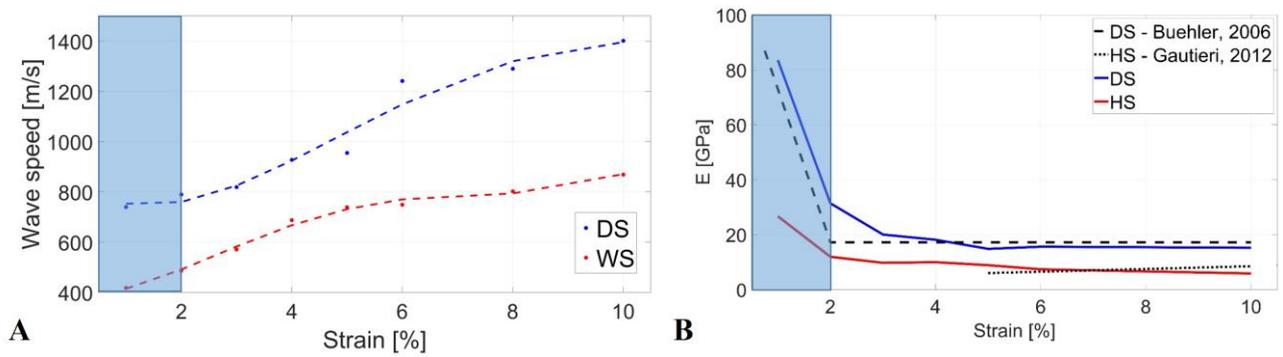

Figure 5 the Wave Speed – Strain and Young's modulus – Strain plots for both the DS and HS. According to the vibrating string model, the wave speed increases monotonically with the strain reaching values of 1401 m·s$^{-1}$ and 869 m·s$^{-1}$ for the DS and HS respectively (

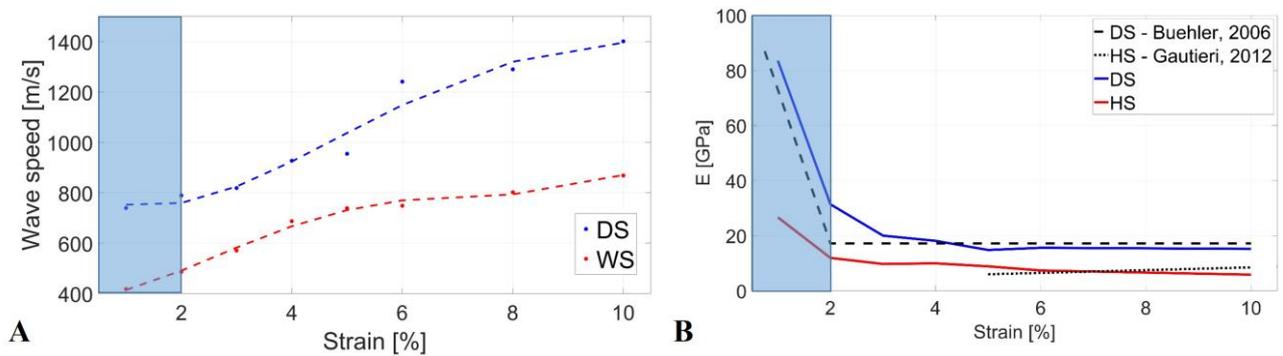

Figure 5A). Compared to the LC, velocities are reduced by a factor of about 2.3, although the highest values are still possessed by the DS. Also in this case, the absence of mineralization reduces the wave speeds and the specific acoustic impedances and the achieved results are definitely smaller than the values estimated for bone when it is exposed to shear waves (2800 m·s$^{-1}$ and 5.32 × 10$^6$ kg·m$^2$·s$^{-1}$ [54]).

To confirm our calculations, we estimate the Young's modulus at different tensile strains and compare the values with previous studies (

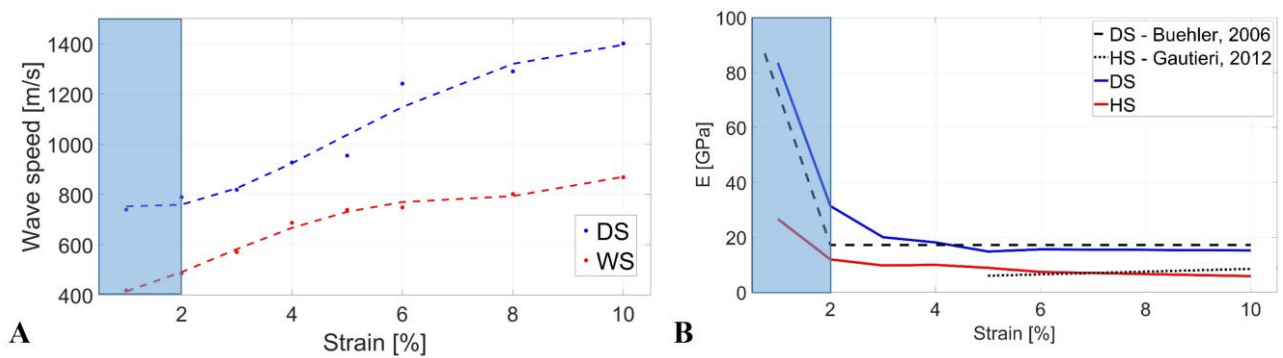

Figure **5**B). For both the structures we achieve a steep decrease of the Young's modulus from very high values at very little strains (83.54 GPa and 26.66 GPa for the DS and HS, respectively), probably due to the bigger errors in estimating the positions of the atoms at small deformations, towards a convergence at 15.21 GPa and 5.85 GPa at 10% of tensile strain, according to earlier studies.[21,26]

## 4. Conclusions

We performed a theoretical study to assess how dry and hydrated collagen peptides are able to transfer and dissipate mechanical energy induced by external impulsive loads. We investigated two different perspectives of the same problem by focusing on the material behavior when applying longitudinal and transversal loads with respect to the triple-helix axis. Concerning the LC, the energy dissipation along the peptide shows similar results between the DS and HS ($\tau_{LC} \approx 100$ ps) whilst the propagation is, as expected, driven by the different stiffnesses that are higher for the DS, according to previous studies.[21] The TC presents a more complex behavior: we found a monotonic increase of the wave speed with the tensile strain, also according to the 1-D vibrating string analytical model, confirming also how the DS is stiffer than HS. However, we observed many interesting differences between the two topologies and in contrast to the LC. Firstly, dissipation behaviors are markedly different for the two peptides, and also in contrast to the LC, with relaxation times $\tau_{DS\_TC} \approx 5 \cdot \tau_{WS\_TC}$ and $\tau_{DS\_TC} \approx 4 \cdot \tau_{LC}$. Concerning the TC, water improves the dissipating effect, rapidly decreasing the amplitude of the displacement waves. This damping effect is clearly absent in the DS, where the deformation is not contrasted by any dissipating agent.

According to best of the authors' knowledge, our study offers a new approach to investigate collagen at the molecular scale, giving new insights on how the triple helix faces external transient loads in its dry or hydrated states. Although employed to study energy propagation and dissipation in proteins,[41] transient loads have not been employed to investigate the behavior of collagen-based materials so far, but they represent a key approach to model thermal management in both healthy and diseased tissues.[56,57]

Hydration level of tissues can vary, reaching also 62% in tendon:[36] therefore, while the HS can well describe structures such as the tendon native state, other tissues may be studied half-way between the hydrated and dry models here investigated. This is the case of the eardrum, a thin spoon-shaped membrane in the middle ear that collects, filter and transfer the transient acoustic inputs to the inner ear. This structure is mainly composed of collagenous fibers radially, circumferentially, and parabolically organized in three different layers.[45] Researchers have studied the structural macro behavior only putting only a limited effort on explaining the mechanical response at fibril and tropocollagen scales. Briefly, upon transient loads, each part of the eardrum vibrates with different frequencies and amplitudes. However, the umbo – the connection with the ossicular chain – receives a filtered energetic content that represents with high fidelity the external input.[40,44,46,58] Since the transversal vibration is the main mechanical response upon transient loads, we believe that there is a tight connection between the macroscale behavior and our study at the molecular level: the orientation of the fibers and, thus, tropocollagen, may contribute to damping the vibration along the fibers promoting, at the same time, the transversal displacement (Figure 6).

Following the interesting results achieved, we plan to apply our method not only to the specific case of pure collagen, but also to the mineralized collagen that represents the building block of the bone,[1] investigating how the covalent cross-links between helixes and different percentages of mineral affect the material behavior. Furthermore, similar studies may be pursued on damaged or diseased tissues by taking into account cases such as *osteogenesis imperfecta* (brittle bone disease) or other specific severe pathologies involving collagen-based tissues (e.g., fibrosis, Ehlers-Danlos syndrome).[59–61] Understanding the energy propagation and dissipation in both healthy and diseased collagen-based tissues at the molecular scale may also lead towards a better understanding of mechanical response of such structures upon transient loads at the macro-scale (e.g., fatigue, impacts).

Furthermore, tissue engineering can benefit from such advancements to develop new biomimetic and bioinspired materials to customize patient-specific devices to replace native tissues.[5,62].In view of this, a remarkable field of application is certainly the middle-ear prosthetics, including ossicular bones and tympanic membrane replacements,[62–70] in which all the structures possess abundant collagen and perform upon acoustic transient loads, to deliver more biocompatible collagen-based devices for long-term outcomes.

**Acknowledgments**

This work was supported by the European Union's Horizon 2020 research and innovation program under the Marie Skłodowska-Curie grant agreement COLLHEAR No 794614. G.S.J. and M.J.B. acknowledges additional support from ONR (N000141612333) and AFOSR (FATE MURI FA9550-15-1-0514), as well as NIH U01HH4977, U01EB014976, and U01EB016422.

The authors want to acknowledge Prof. Elizabeth S. Olson (Columbia University, New York, NY) for fruitful discussions and suggestions.

**Conflict of interest**

All the authors declare no conflict of interest.

**Authors contributions**

Conceptualization, M.M., S.D., and M.J.B.; Data curation, M.M. and G.S.J.; Formal analysis, M.M. and G.S.J.; Funding acquisition, M.M., M.J.B.; Investigation, M.M., S.D., and M.J.B.; methodology; Project administration, M.M., M.J.B.; Resources, M.J.B.; Software, M.J.B.; Supervision, S.D., and M.J.B.; Validation, M.M. and G.S.J.; Visualization, M.M.; Roles/Writing - original draft, M.M.; Writing - review & editing, M.M., G.S.J, S.D., and M.J.B..

All authors intellectually contributed and provided approval for publication.

**Abbreviations:** Hydrogen bonds, H-Bonds; DS, Dry Structure; HS, Hydrated Structure; LC, Longitudinal Case; TC, Transversal Case.

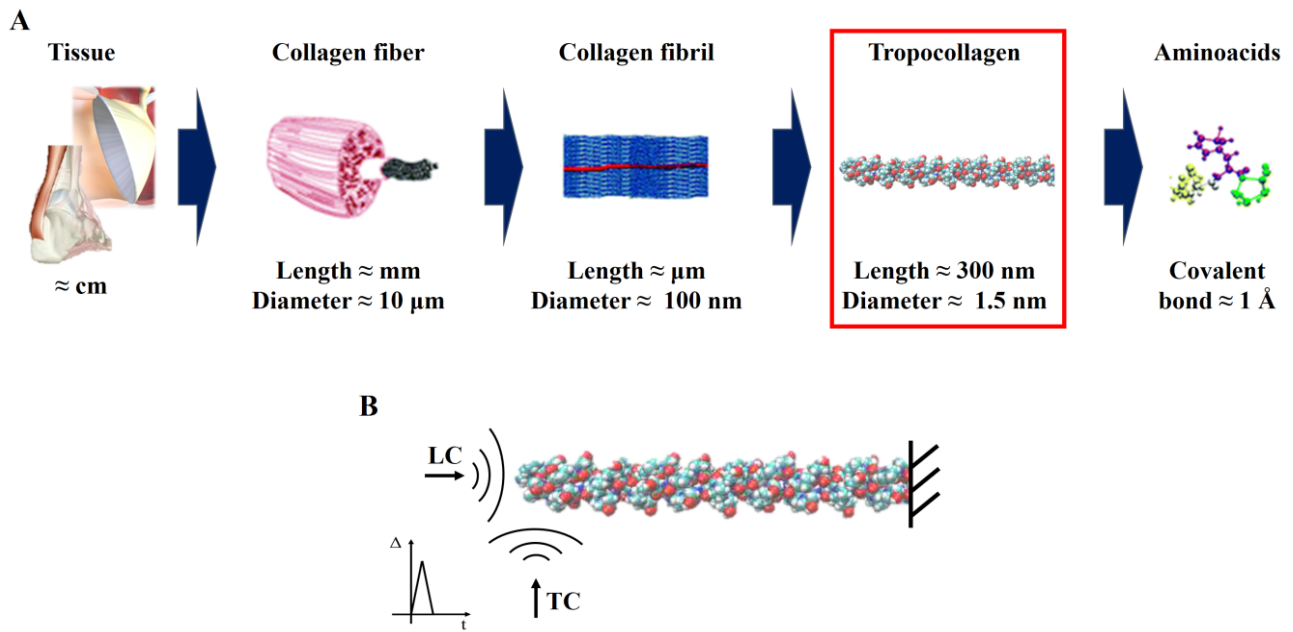

*Figure 1.* Panel A. Hierarchical structure of collagenous tissues with characteristic length scales: from macro tissues (e.g., tendon, eardrum) to amino acids [7–15]. Tropocollagen (red lens) is the subject of this study. Panel B. Aiming at investigating wave propagation and energy dissipation along the triple helix, we use longitudinal (Longitudinal Case – LC) and transversal (Transversal Case – TC) impulsive loads at the free end of a dry (Dry Structure – DS) and hydrated (Hydrated Structure – HS) tropocollagen segment (length ≈ 180 Å). Adapted with permission from [25]. Copyright 2011 American Chemical Society.

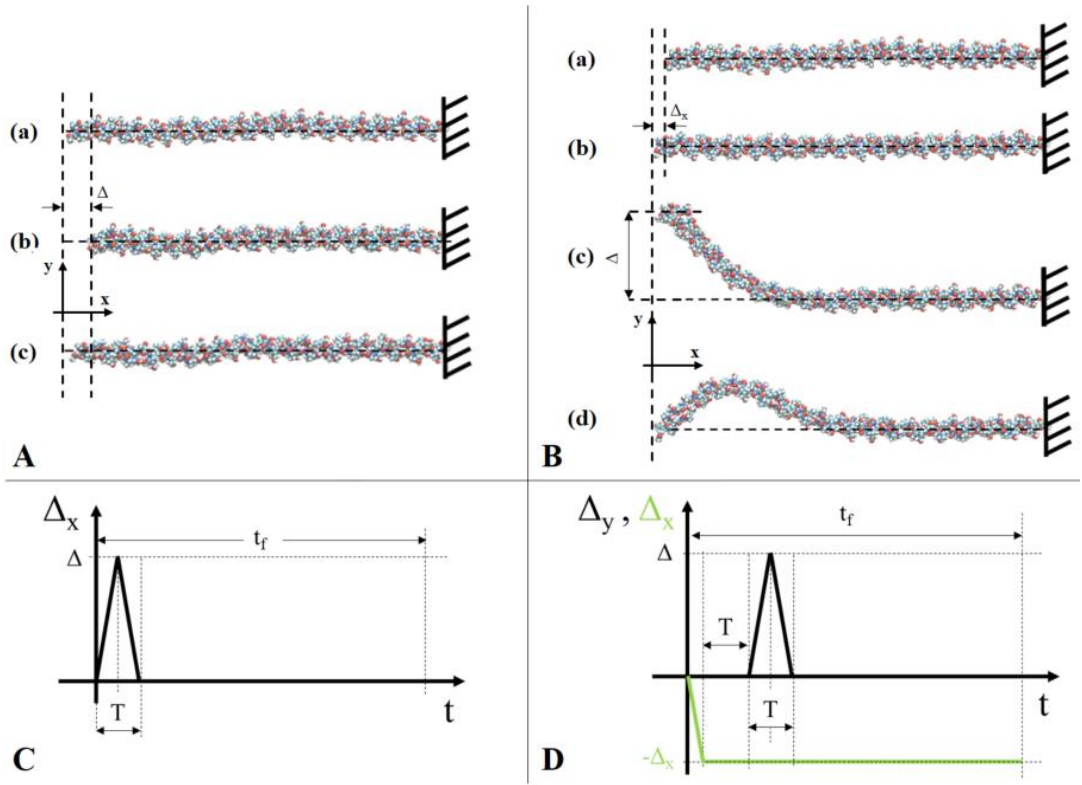

***Figure 2***. *Collagen peptides upon impulsive loads and fixed at one end. The impulsive load has amplitude Δ equals 10 Å, a slope of 100 m·s$^{-1}$, and period T equals 10 ps. The global time of observation is $t_f$ equals 160 ps. LC – Panel A: the collagen peptide is loaded axially with displacement $Δ_x$ as shown in panel C. TC – Panel B: the collagen peptide is firstly axially stretched with a slope of 100 m·s$^{-1}$ up to a set tensile strain (i.e., from 1 to 10%) and (b) deformed from the free end with a vertical impulsive load Δ after 10 ps, as shown in panel D. For both cases, (a) represents the relaxed state, (b) – and (c) for the TC – shows a loading frame, and (c) – (d) for the TC – depicts a frame after the load application.*

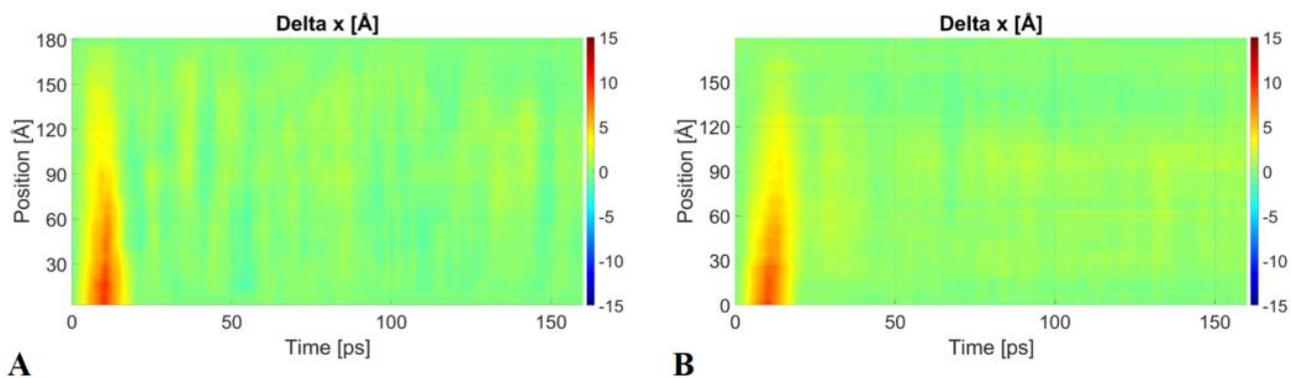

*Figure 3*. Plots of DS (A) and HS (B) $C_\alpha$ axial displacement [Å] over time and position along the peptide. In both cases, note the significant energy dissipation that annihilates almost completely the wave already before it reaches the fixed edge. DS shows a steeper slope of the traveling displacement leading to a higher velocity of the mechanical wave and, thus, of the Young's modulus, according to [21].

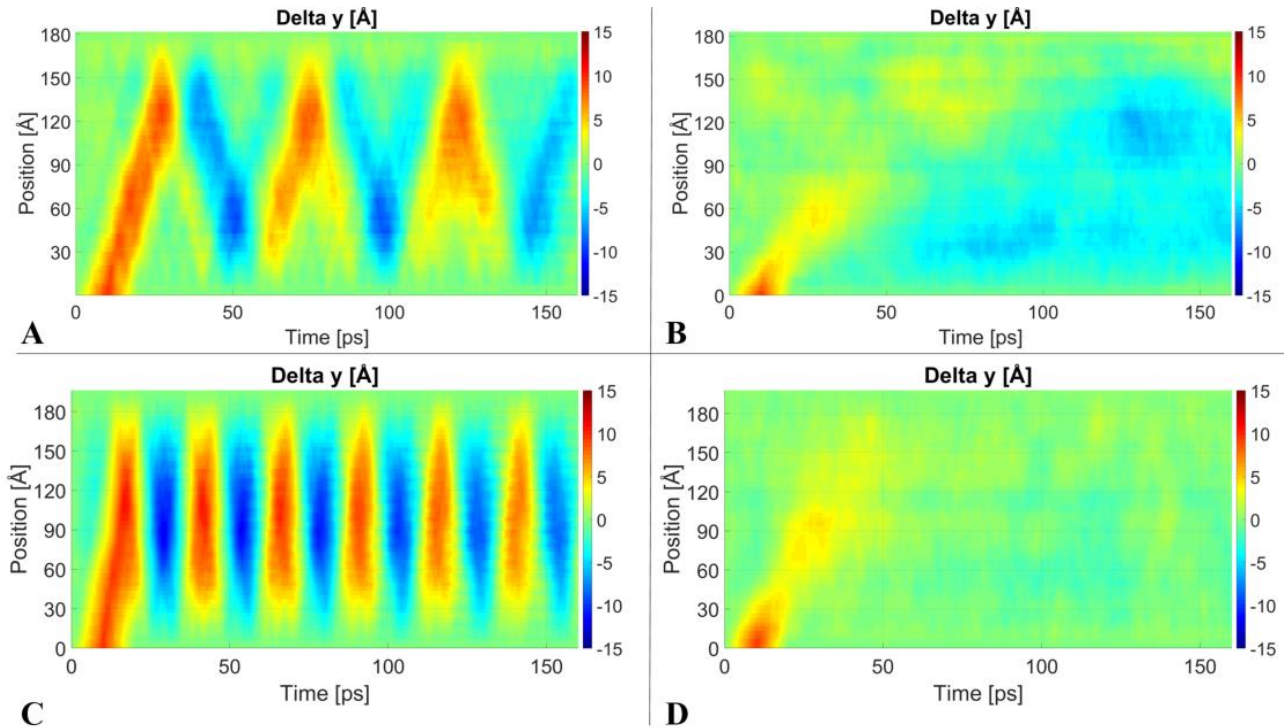

*Figure 4*. TC – Transversal displacement [Å] along the peptide for dry (A, C) and hydrated (B, D) collagen peptides as a function of time, following a preliminary axial displacement of 1 Å (A, B) and 18 Å (C, D). Contrary to the LC, dissipation is significant only for the wet peptide. In contrast, dry collagen presents a more elastic, less viscous, behavior. By modeling the peptide as a pre-tensioned string, the most pulled structure possesses the highest wave speed, here depicted with steeper slopes of the traveling displacement. Finally, dry cases show higher velocities reflecting higher Young's moduli.

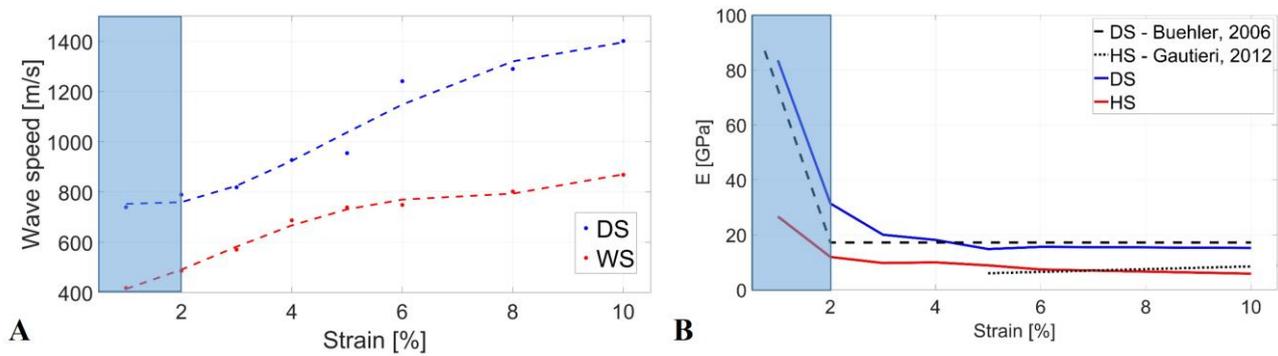

*Figure 5*. *TC – Depiction of the Wave speed – Strain (A) and Young's modulus – Strain (B) curves for the DS and HS. As also predictable with the vibrating string model, Panel A shows how the wave speed along the peptide increases monotonically with the pre-strain. Panel B, as a consequence of Panel A, shows a comparison between the Young's modulus estimated through the wave propagation and previous results achieved through slowly stretching either DS or HS. Interestingly, the blue curve, representing our results for DS, well matches with the outcomes from* [26]. *Similarly, the estimated Young's modulus for HS fits the available data from* [21]. *The high jump at low strains could be explained by the highest error in estimating the Young's modulus in such small deformations. Therefore, we exclude from our discussion the results highlighted with the blue rectangle in both the Panels.*

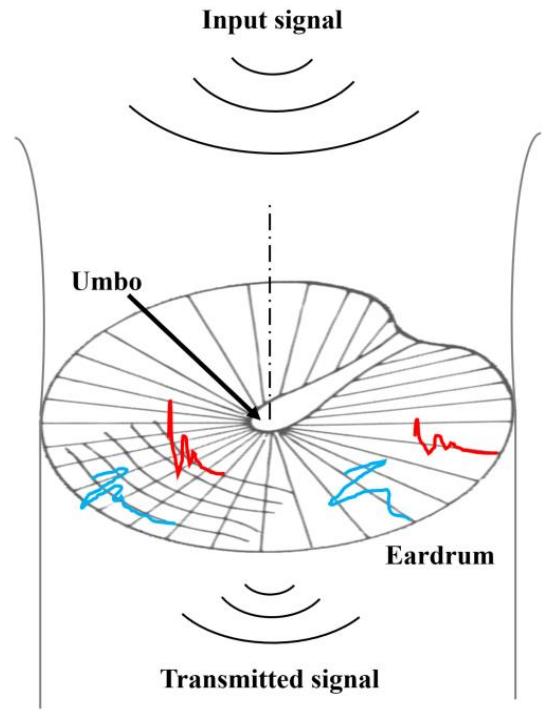

*Figure 6. Input signal filtering and dissipation by the eardrum. The tympanic membrane, with its collagenous macro-structured radial, circumferential, and parabolic fibers, is able to deliver to the umbo – the connection with the ossicular chain – only specific energy contents despite its complex vibrating behavior* [40,44,58]. *We believe that the mechanisms studied in this paper may be correlated to the mechanical response of the eardrum, which privileges transversal vibrations (red curves) damping more the longitudinal displacements (blue curves) along the fibers.*

# Wave propagation and energy dissipation in collagen molecules

Mario Milazzo, Gang Seob Jung, Serena Danti, Markus J. Buehler

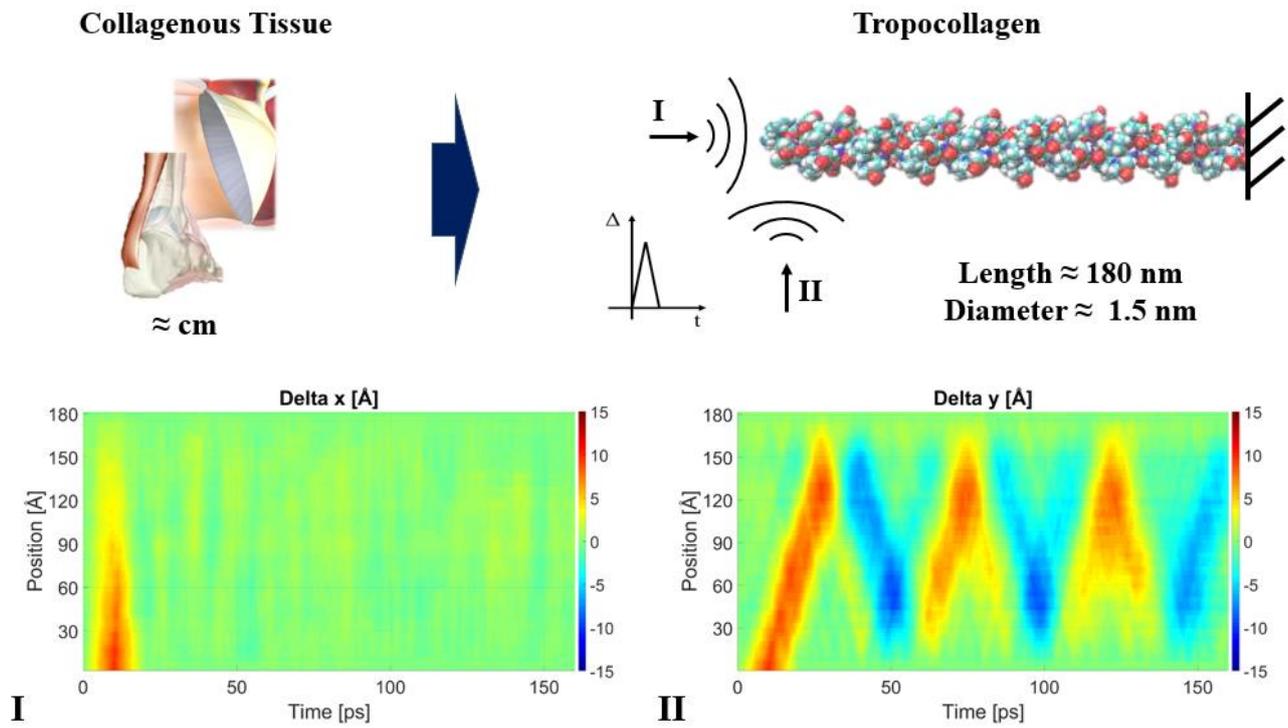

***TOC image***: *Wave propagation and energy dissipation of collagen molecules.*